# Extending spin dephasing time of perfectly aligned Nitrogen-Vacancy centers by mitigating stress distribution on highly misoriented chemical-vapor-deposition diamond

T. Tsuji[1], T. Sekiguchi[1], T.Iwasaki[1] and M.Hatano[1*]

[1] *Department of Electrical and Electronic Engineering, School of Engineering,*

*Tokyo Institute of Technology, 2-12-1 Ookayama, Meguro, Tokyo 152-8552, Japan*

E-mail: hatano.m.ab@m.titech.ac.jp

Abstract

Extending the spin-dephasing time ($T_2^*$) of perfectly aligned nitrogen-vacancy (NV) centers in large-volume chemical vapor deposition (CVD) diamonds leads to enhanced DC magnetic sensitivity. However, $T_2^*$ of the NV centers is significantly reduced by the stress distribution in the diamond film as its thickness increases. To overcome this issue, we developed a method to mitigate the stress distribution in the CVD diamond films, leading to a $T_2^*$ extension of the ensemble NV centers. CVD diamond films of approximately 50 μm thickness with perfectly aligned NV centers were formed on (111) diamond substrates with misorientation angles of 2.0, 3.7, 5.0, and 10°. We found that $T_2^*$ of the ensemble of NV centers increased to approach the value limited only by the electron and nuclear spin bath with increasing the misorientation angle. Microscopic stress measurements revealed that the stress distribution was highly inhomogeneous along the depth direction in the CVD diamond film at low misorientation angles, whereas the inhomogeneity was largely suppressed on highly misoriented substrates. The reduced stress distribution possibly originates from the reduction of the dislocation density in the CVD diamond. This study provides an important method for synthesizing high-quality diamond materials for use in highly sensitive quantum sensors.

# 1. Introduction

Negatively charged nitrogen-vacancy (NV) centers in diamonds work as room-temperature quantum sensors. NV centers are sensitive to a wide range of physical parameters, such as the magnetic field[1,2], electric field[3,4], temperature[5,6], and pressure[7,8], and are used in applications with spatial resolutions from the nanoscale [9,10] to the millimeter scale [11–13]. In particular, sensors with large detection volumes (>(0.1 mm)$^3$) are expected to achieve high magnetic sensitivity required for biomagnetic measurements [14–16] and battery current monitoring[17].

The DC magnetic sensitivity of the ensemble NV centers can be enhanced by improving the measurement contrast ($C$), number of NV centers ($N$), and spin dephasing time ($T_2^*$) [18]. $C$ can be improved by controlling the alignment of the NV centers. The perfectly aligned NV centers formed by the chemical vapor deposition (CVD) method [19–23] can achieve four times higher $C$ than the NV centers formed by electron-beam irradiation and ion implantation methods, where the NV centers are randomly aligned in four directions: ([111], [$\bar{1}\bar{1}$1], [1$\bar{1}\bar{1}$], and [$\bar{1}$1$\bar{1}$]). To form perfectly aligned NV centers, the diamond needs to be grown by step-flow growth [24] on a (111) diamond substrate polished along [$\bar{1}\bar{1}$2] direction at a misorientation angle ($\theta_{mis}$). Increasing the volume (thickness) of the diamond film containing the NV centers improves $N$. We have so far reported the formation of a thick diamond film (>100 µm) with perfectly aligned NV centers [23]. The improvement of the third parameter, the spin dephasing time $T_2^*$, remains a challenge on the CVD formation of aligned NV centers.

The $T_2^*$ of ensemble NV centers is mainly reduced by electron spin bath, such as P1 centers, nuclear spin bath of $^{13}$C, and stress distribution in the diamond lattice[18]. The reduction in $T_2^*$ due to the spin baths is caused by the dipolar interaction between the NV centers and these spin baths[25]. A strain field shifts the resonance frequencies of an NV center by spin-stress interaction [26,27]. Therefore, an inhomogeneous stress distribution in a thick diamond film causes a significant broadening of the electron spin resonance signal from the ensemble NV centers, resulting in the $T_2^*$ reduction [18].

In this study, we found and established a method for improving $T_2^*$ by mitigating the inhomogeneous stress distribution in the thick diamond film with perfectly aligned NV centers. CVD diamond films of approximately 50 µm thickness with perfectly aligned NV centers were synthesized on (111) diamonds with misorientation angles ($\theta_{mis}$) of 2.0, 3.7, 5.0, and 10°. The $T_2^*$ of the large-ensemble NV centers was evaluated using a measurement system in which an excitation laser illuminated the whole thickness of the CVD film. We found that the $T_2^*$ of large-ensemble NV centers increased to approach its value, limited only by the spin bath, with increasing $\theta_{mis}$. A confocal microscope was used for microscopic spatial imaging of the stress distribution in the diamond film. The stress was highly inhomogeneous along the depth direction in the CVD diamond film, and the stress distribution decreased with increasing $\theta_{mis}$. These results demonstrate that $T_2^*$ of the large-ensemble NV centers increased by mitigating the stress distribution. The suppression of dislocation formation in the CVD diamond for a high $\theta_{mis}$ is considered to contribute to the reduced stress distribution.

## 2. Material and methods

Type-Ib HPHT (111) diamonds were used as substrates for homo-epitaxial diamond growth. The surfaces of the (111) diamond substrates were polished along $[\bar{1}\bar{1}2]$ direction at $\theta_{mis}$ of 2.0, 3.7, 5.0, and 10°, as shown in Fig.1 (a), which were measured using X-ray diffraction. We used a high-power microwave plasma CVD with a spherical chamber so that the microwaves reflectively concentrated on the diamond substrate[22]. Hydrogen, methane, and nitrogen gases flowed at 500, 0.5, and 0.01 sccm, respectively. The pressure, microwave power, and temperature were set to 30 kPa, 2.1 kW, and 800 °C, respectively. The growth time was set so that the CVD film thickness becomes approximately 50 μm, considering that the growth rate varies with the $\theta_{mis}$ [23]. Fig.1 (b) shows the optical image of the CVD diamond for $\theta_{mis}$ of 10°. Fig.1 (c) shows a cross-sectional schematic of the P-S line shown in Fig.1 (b). We confirmed that the diamonds were grown by step-flow growth along $[\bar{1}\bar{1}2]$ direction for all $\theta_{mis}$. We measured $T_2^*$ and the stress distribution on the flat surface between the Q-R lines for all $\theta_{mis}$.

The spin dephasing time $T_2^*$ of large ensemble NV centers was evaluated with a measurement system in which a 532 nm laser was focused to a diameter of 20 μm on the CVD film to excite the whole thickness of the CVD film, as shown in Fig.1 (d) [28]. A coplanar waveguide resonator was used for a strong and uniform microwave irradiation perpendicular to the NV axis along the [111] direction[29]. A ring magnet was used to apply a uniform magnetic field of approximately 7 mT to the whole of the excitation volume. (Details are provided in the Supporting Information). This measurement system is referred to as a large-excitation-volume system in this study.

A confocal microscope was used to map the stress distribution within a volume set at the same location as that used in the large-excitation-volume system, as shown in Fig.1 (e). A diamond sample was placed on a piezoelectric stage, and a permanent magnet was used to apply a magnetic field in the [111] direction (details are provided in the Supporting Information). The stress distribution was measured on the cross-sectional plane and surface of the CVD film. The deviation of the z component of the spin-stress interaction ($\Delta M_z$), which is the major cause of the reduction in $T_2^*$ described below, was measured at 6×6 = 36 positions on the cross-sectional plane and 6×6 = 36 positions on the surface of the CVD film. The optically detected magnetic resonance (ODMR) spectra of the NV centers were measured using a confocal microscope under a static magnetic field along the [111] direction.

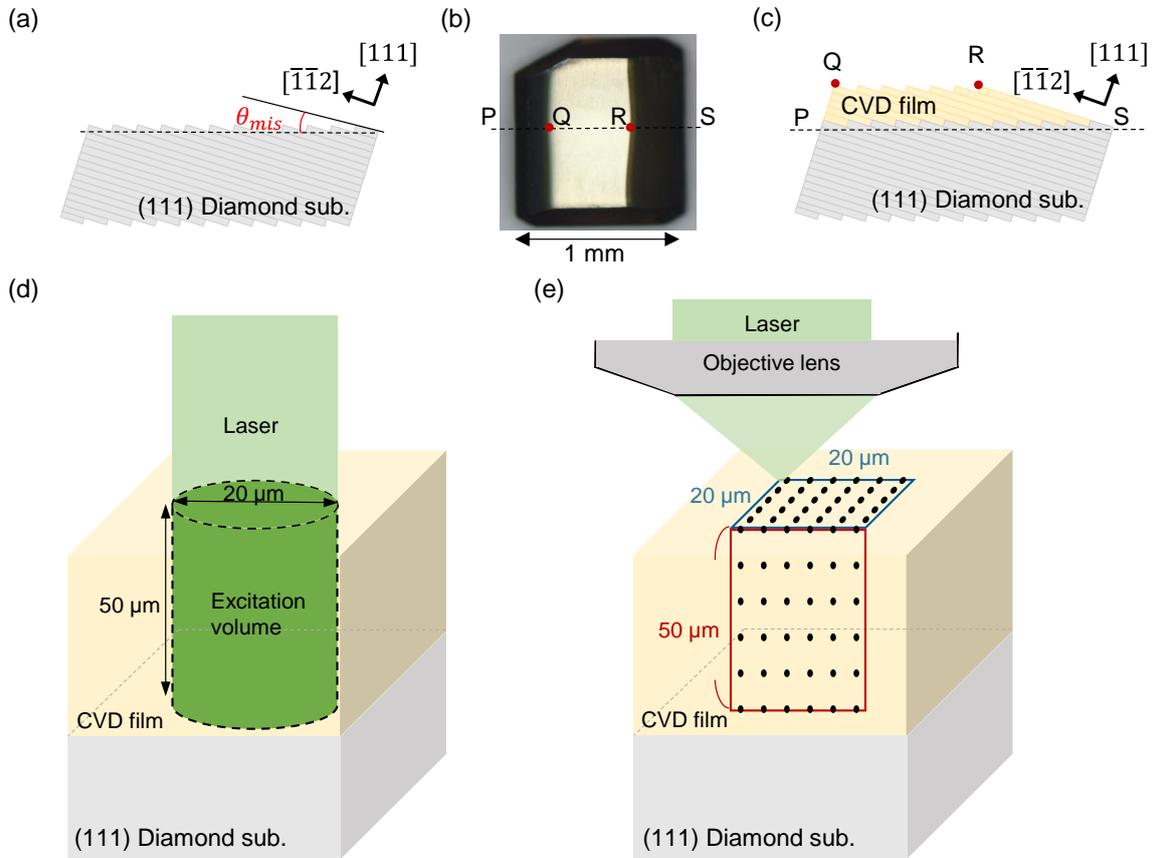

Fig.1 (Color online) (a) Misorientation angle $\theta_{mis}$ defined as the angle between the polished diamond substrate plane and $[\bar{1}\bar{1}2]$ direction. (b) Optical image of the CVD diamond for $\theta_{mis}$ of 10°. (c) Cross-sectional schematic image of a P-S line shown in Fig.1 (d) Schematic of the large excitation volume system where the excitation laser focused to a diameter of 20 μm illuminated on the CVD film to excite the whole thickness of the CVD film. (e) Schematic of a confocal microscope for measuring stress distribution. The stress distribution was measured within a volume, which was set at the same location as used in the large excitation volume system. The deviation of the z component of spin-stress interaction ($\Delta M_z$) was measured at 6×6 = 36 positions on the cross-section plane and 6×6 = 36 positions on the surface in the CVD film.

## 3. Results and discussion

Fig.2 (a) shows the ODMR spectrum of the NV centers for $\theta_{mis} = 10°$ obtained using a confocal microscope. Only two dips were observed for all $\theta_{mis}$, confirming the reproducible formation of perfectly aligned NV centers for all $\theta_{mis}$ as reported in a previous study [23].

We evaluated $T_2^*$ of large-ensemble NV centers in the CVD films using a large-excitation-volume system. Fig.2 (b) shows the Ramsey fringes of the NV centers at $\theta_{mis}$ of 10° measured by the large excitation volume system. $T_2^*$ was extracted from the fitting to the function of $\exp\left(-\frac{\tau}{T_2^*}\right)[a_1 \cos(2\pi(f-f_h)\tau + \varphi_1) + a_2 \cos(2\pi f \tau + \varphi_2) + a_3 \cos(2\pi(f+f_h)\tau + \varphi_3)]$, where $f_h$ (=2.177 MHz) represents the hyperfine splitting. Fig.2 (c) shows the dependence of $T_2^*$ measured using a large-excitation-volume system on $\theta_{mis}$. We found that $T_2^*$ increased from approximately 0.15 μs to 0.30 μs as $\theta_{mis}$ increased from 2.0° to 5.0° and saturated at approximately 0.30 μs for $\theta_{mis} \geq 5.0°$.

$T_2^*$ measured using a large excitation volume system ($T_2^*\{\text{Measure}\}$) can be decomposed into contributions from different dephasing sources as follows:

$$\frac{1}{T_2^*\{\text{Measure}\}} = \frac{1}{T_2^*\{\text{Electron spin}\}} + \frac{1}{T_2^*\{\text{Nuclear spin}\}} + \frac{1}{T_2^*\{\text{Stress}\}} + \frac{1}{T_2^*\{\text{Other}\}}, \quad (1)$$

where $T_2^*\{\text{Electron spin}\}$, $T_2^*\{\text{Nuclear spin}\}$, $T_2^*\{\text{Stress}\}$, and $T_2^*\{\text{Other}\}$ are the hypothetical limits to $T_2^*$ solely due to electron spins, such as the P1 center ($N_s^0$), nuclear spins, such as $^{13}C$, stress distribution in the CVD film, and other factors [18], respectively. $T_2^*\{\text{Electron spin}\}$ was assumed to depend mainly on the P1 center density ($[N_s^0]$) according to the following equation:

$$T_2^*\{\text{Electron spin}\} = \frac{\alpha}{[N_s^0]}, \quad (2)$$

where $\alpha$=9.6 ± 0.9 μs×ppm[25]. $[N_s^0]$ in the CVD film was calculated from the spin coherence time ($T_2$) measured by the Hahn echo sequence using the confocal microscope as follows:

$$[N_s^0] = \frac{\beta}{T_2}, \quad (3)$$

where $\beta$=160 ± 12 μs×ppm[25,30]. $T_2$ was measured at 3 μm intervals in the depth direction in CVD film for each $\theta_{mis}$. The $T_2$ values used to calculate $[N_s^0]$ and $T_2^*\{\text{Electron spin}\}$ calculated using Eq. (2) are presented in Table 1 (The Details of the results are provided in the Supporting Information.) The CVD diamond films formed in this study contained 1.1% of $^{13}$C nuclear spin bath because the methane gas used in this study had a natural abundance representing $T_2^*\{\text{Nuclear spin}\} \approx 1.0$ MHz [18]. Thus, $1/T_2^*\{\text{Stress}\} + 1/T_2^*\{\text{Other}\}$ can be evaluated by $\frac{1}{T_2^*\{\text{Stress}\}} + \frac{1}{T_2^*\{\text{Other}\}} = \frac{1}{T_2^*\{\text{Measure}\}} - \frac{1}{T_2^*\{\text{Electron spin}\}} - \frac{1}{T_2^*\{\text{Nuclear spin}\}}$.

Fig.2 (d) shows the $\theta_{mis}$ dependence of each component of $1/T_2^*$. As $\theta_{mis}$ was increased, $1/T_2^*\{\text{Measure}\}$ decreased significantly and $1/T_2^*\{\text{Electron spin}\} + 1/T_2^*\{\text{Nuclear spin}\}$ was almost unchanged as explained above. $1/T_2^*\{\text{Measure}\}$ approached the hypothetical limit of $T_2^*$ due to electron spins and nuclear spins ( $1/T_2^*\{\text{Electron spin}\} + 1/T_2^*\{\text{Nuclear spin}\}$ ) as $\theta_{mis}$ increased. In other words, $1/T_2^*\{\text{Stress}\} + 1/T_2^*\{\text{Other}\}$ decreased with increasing $\theta_{mis}$ and almost vanished for $\theta_{mis} \geq 5.0°$. More quantitatively, $1/T_2^*\{\text{Stress}\} + 1/T_2^*\{\text{Other}\}$ decreased by a factor of at least 19 as $\theta_{mis}$ was increased from 2.0° to 10°. $1/T_2^*\{\text{Other}\}$ includes electric field noise, temperature fluctuations, and magnetic field gradients [18]. All these contributions were considered constant for all $\theta_{mis}$ because the same measurement conditions were used for the evaluation of $T_2^*\{\text{Measure}\}$ for all $\theta_{mis}$. Therefore, the $1/T_2^*\{\text{Other}\}$ were negligible in these measurements. Furthermore, it suggests that the decrease in $1/T_2^*\{\text{Stress}\}$ should have a dominant contribution to the observed decrease in $1/T_2^*\{\text{Stress}\} + 1/T_2^*\{\text{Other}\}$.

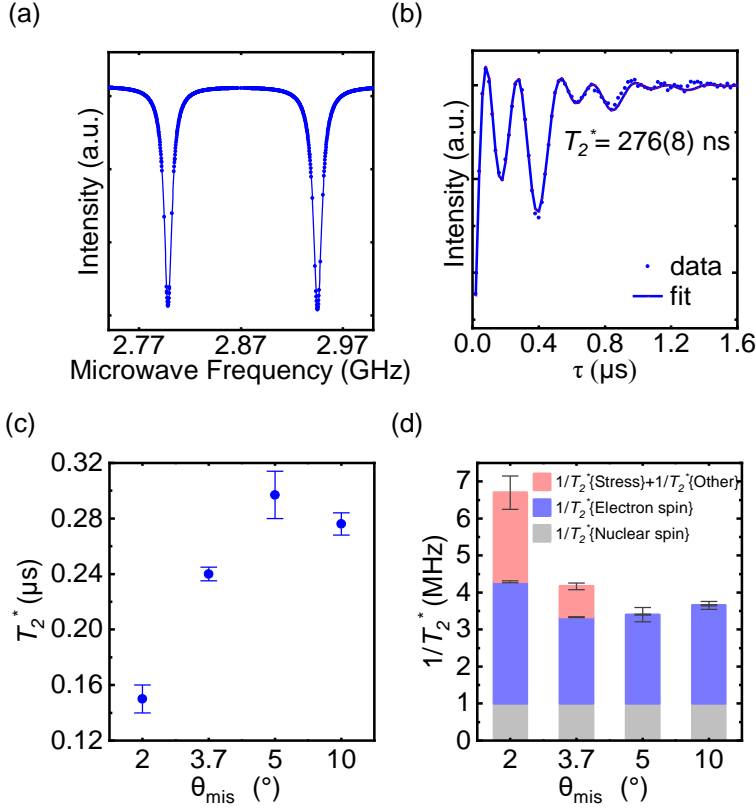

Fig.2 (Color online) (a) ODMR spectrum of NV centers at $\theta_{mis}$ of 10° measured with the confocal microscope. (b) Ramsey fringes of NV centers at $\theta_{mis}$ of 10° measured with the large detection volume system. (c) Dependence of $T_2^*$ on $\theta_{mis}$ measured with the large excitation volume system. (d) Dependence of each component of $1/T_2^*${Electron spin}, $1/T_2^*${Nuclear spin}, and $1/T_2^*${Stress} + $1/T_2^*${Other}, on $\theta_{mis}$.

Table 1: Average value of $T_2$, $[N_s^0]$ and $T_2^*${Electron spin} for each $\theta_{mis}$. $[N_s^0]$ and $T_2^*${Electron spin} were calculated using Eqs. (3) and (2), respectively.

| $\theta_{mis}$ (°) | $T_2$ (μs) | $[N_s^0]$ (ppm) | $1/T_2^*[Electron\ spin]$ (MHz) |
|---|---|---|---|
| 2.0 | 5.07(4) | 31.6(2) | 3.29(3) |
| 3.7 | 7.14(4) | 22.4(1) | 2.33(1) |
| 5.0 | 6.94(3) | 23.0(1) | 2.40(1) |
| 10 | 6.28(4) | 25.5(2) | 2.65(2) |

Next, the stress distribution in the CVD film was measured using confocal microscopy, as shown in Fig.1 (e). Fig.3 (a) shows the fluorescence images of the surface (XY plane) and cross-section (XZ plane) using the confocal microscope. The deviation of the z component of spin-stress interaction ($\Delta M_z$), which is the major cause of the reduction in $T_2^*$ as described below, was measured at 6×6 = 36 positions on the XZ plane and 6×6 = 36 positions on the XY plane. In Fig.3 (a), Z indicates the distance moved by the stage. Considering the refractive index difference between the immersion oil (1.52) and diamond (2.42), the thickness of the diamond film (≈ 50 μm) is different from 20 μm in the range of Z.

The NV center's electronic ground state Hamiltonian in the presence of stress and a static magnetic field is [31]:

$$H = (D + M_z)S_z^2 + \gamma \vec{B} \cdot \vec{S} - M_X(S_X^2 - S_Y^2) + M_Y(S_X S_Y + S_Y S_X), \tag{4}$$

where $D \approx$ 2.87GHz is the temperature-dependent zero-field splitting parameter, $\gamma$=28.03 GHz/T is the NV gyromagnetic ratio, $\vec{S} = (S_X, S_Y, S_Z)$ are the spin-1 operators, $\vec{B}$ is the applied magnetic field, and $\vec{M} = (M_X, M_Y, M_Z)$ is spin-stress interaction. Using the second-order perturbation theory, the resonance frequencies of the NV center ($f_\pm$) can be expressed as

$$f_\pm = E_{\pm 1} - E_0 = D + M_z \pm \gamma B_z + \frac{\gamma^2(B_x^2 + B_y^2)}{D + M_z \pm \gamma B_z} + \frac{\gamma^2(B_x^2 + B_y^2)}{2(D + M_z \mp \gamma B_z)} \pm \frac{M_x^2 + M_y^2}{2\gamma B_z}, \tag{5}$$

where $E_{m_s}$ ($m_s = +1, 0, -1$) are the energy levels of the NV center. From equations (5), the broadening of the resonance frequencies of ensemble NV centers, that is, reduction of $T_2^*$ due to spin-stress interaction ($M_X, M_Y, M_Z$) are caused by the distribution of $M_z$ and $\frac{M_x^2 + M_y^2}{\gamma B_z}$. Here, we confirmed the validity of $|M_z| \gg \left|\frac{M_x^2 + M_y^2}{\gamma B_z}\right|$ since we applied the magnetic field of $B_z \approx$ 7 mT in the large excitation volume system and spin-stress interaction were typically less than 2 MHz as described below (Figs.3 (b) and (c)). Therefore, the major cause of the reduction in $T_2^*$ through the spin-stress interaction is the distribution of $M_z$.

We defined $\alpha(P)$ as

$$\alpha(P) = (f_-(P) + f_+(P))/2, \tag{6}$$

where $P$ is each measurement position shown in Fig.3 (a). We can calculate the deviation of $\alpha(P)$ as

$$\alpha(P) - \alpha(P_0) \approx \Delta M_z(P) \tag{7}$$

because D≫γB$_z$, D≫$M_z$, and D≫$B_x^2 + B_y^2$, where $P_0$ was the reference position shown in Fig.3 (a) and the deviation of $M_z$ was defined as $M_z(P) - M_z(P_0) = \Delta M_z(P)$ (details are provided in the Supporting Information). The two resonance frequencies ($f_\pm(P)$) of the NV center at each measurement position $P$ were measured using the Ramsey sequence. We applied a microwave of frequency $f_{mw}$ a few megahertz shifted upward from each of the resonance frequencies of the NV center ($f_\pm$) to obtain Ramsey fringes. The Ramsey fringe was fitted using the following function $\exp\left(-\frac{\tau}{T_2^*}\right)[a_1 \cos(2\pi(f_{0\pm} - f_h)\tau + \varphi_1) + a_2 \cos(2\pi f_{0\pm}\tau + \varphi_2) + a_3 \cos(2\pi(f_{0\pm} + f_h)\tau + \varphi_3)]$, where $f_h$ =2.177 MHz represents the hyperfine splitting. Thereby, the resonance frequency of the NV center was accurately obtained by $f_\pm = f_{mw} - f_{0\pm}$ (details are provided in the Supporting Information). Finally, we calculated the standard deviation of the $\Delta M_z(P)$ in the XZ and XY planes to evaluate the distribution of $\Delta M_z$, which was the major cause of the reduction in $T_2^*$.

Fig.3 (b) shows the depth (Z) dependence of $\Delta M_z$ in the cross section (XZ plane) of the CVD film for different $\theta_{mis}$. Each data point represents the mean value of $\Delta M_z$ for six measurement points along the X-direction at a given Z, and each error bar represents the standard deviation of $\Delta M_z$ for these six points. Fig.3 (c) shows the X-dependence of $\Delta M_z$ on the surface (XY plane) of the CVD film for different $\theta_{mis}$. Each data point represents the mean value of $\Delta M_z$ for six measurement points along the Y-direction at a given X, and each error bar represents the standard deviation of $\Delta M_z$ for these six points. $\Delta M_z$ was distributed from approximately -2 MHz to +1 MHz at $\theta_{mis}$ = 2° on the XZ plane. The distribution of $\Delta M_z$ became narrower as $\theta_{mis}$ was increased on the XZ planes. At $\theta_{mis}$ of 10°, the distribution of $\Delta M_z$ was shrunk ranging from approximately -0.12 MHz to +0.02 MHz on the XZ plane. On the other hand, $\Delta M_z$ was distributed from approximately -0.5 MHz to +0.8 MHz at $\theta_{mis}$ of 2° on the XY plane. The distribution of $\Delta M_z$ also decreased as $\theta_{mis}$ increased in the XY plane. At $\theta_{mis}$ of 10°, the distribution of $\Delta M_z$ was shrunk ranging from approximately -0.08 MHz to +0.08 MHz on the XY plane. $\Delta M_z$ had a wider distribution on the XZ plane than on the XY plane for each $\theta_{mis}$. In addition, $\Delta M_z$ monotonically increased as the distance Z from the substrate was increased. Therefore, we considered that the stress accumulated along the distance from the substrate as the film thickness increased. Fig.3 (d) shows the standard deviation of $\Delta M_z$ at 6×6=36 measurement points on the XZ plane and 6×6=36 measurement points on the XY plane for different $\theta_{mis}$. We found that the standard deviation decreased by a factor of approximately 24 in the XZ plane and 11 in the XY plane as $\theta_{mis}$ increased from 2° to 10°. This result indicates that the stress distribution was mitigated by increasing $\theta_{mis}$. Therefore, we conclude that the increase of $\theta_{mis}$ for the CVD growth efficiently extended the $T_2^*$ of the NV centers in the CVD film, as shown in Figs.2 (c) and (d), through strong suppression of inhomogeneous distribution of the $M_z$ stress component.

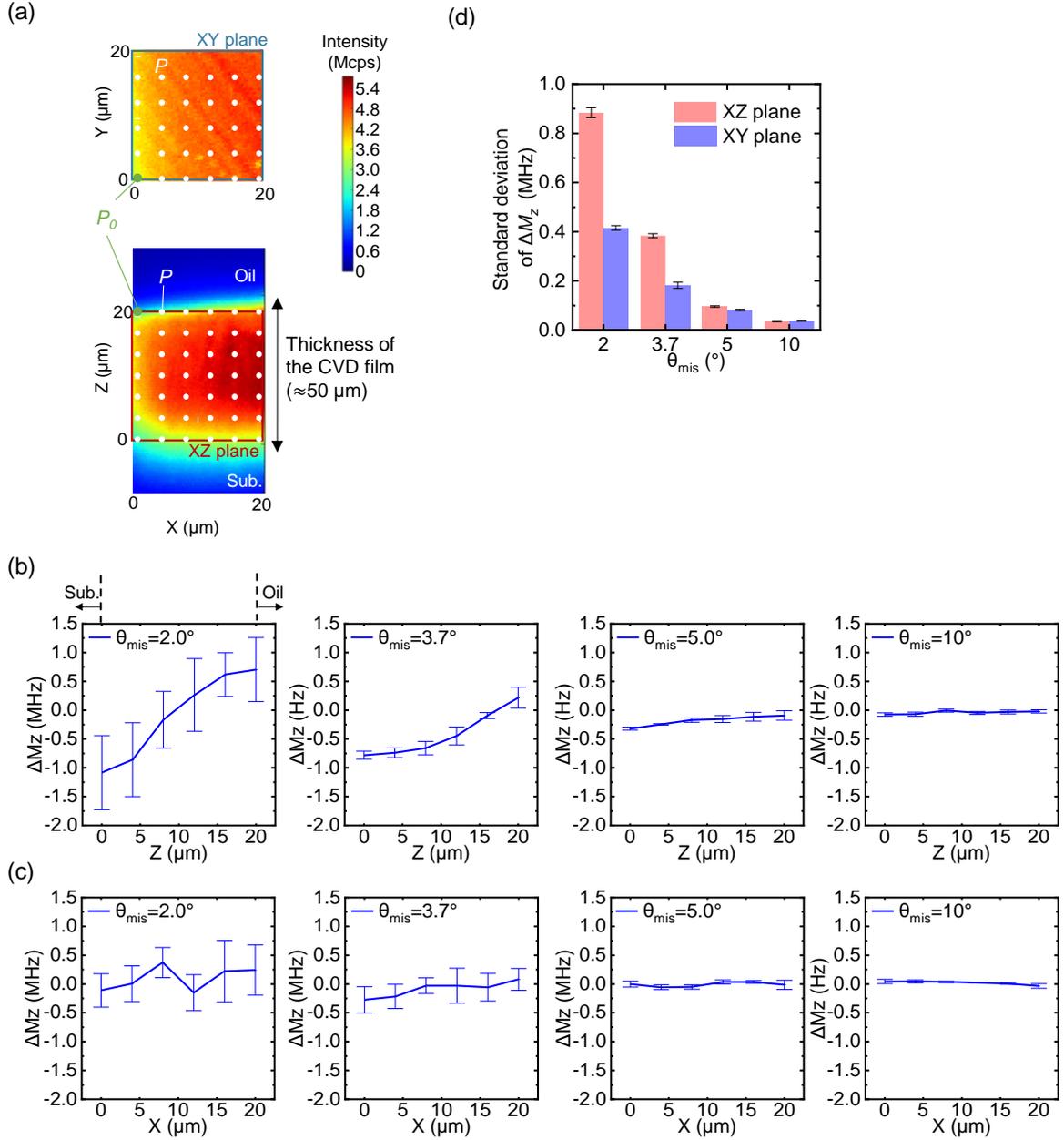

Fig. 3 (Color online) (a) Fluorescence images in the surface (XY plane) and cross-section (XZ plane) of the CVD film using the confocal microscope. A green dot and white dots represent a reference position and measurement positions, respectively. $P$ (white dots) and $P_0$ (a green dot) are each measurement position and a reference position, respectively. Z is the distance the stage has moved. Note that, considering the refractive index difference between oil and diamond, the thickness of the diamond film ($\approx$ 50 μm) is different from 20 μm in the range of Z. (b) Z dependence of $\Delta M_z$ in the XZ plane for different $\theta_{mis}$. Each data point represents the mean value of $\Delta M_z$ for six measurement points along the X direction at a given Z, and each error bars represent the standard deviation of $\Delta M_z$ for these six points. (c) X dependence of $\Delta M_z$ in the XY plane for different $\theta_{mis}$. Each data point represents the mean value of $\Delta M_z$ for six measurement points along the Y direction at a given X, and each error bars represent the standard deviation of $\Delta M_z$ for these six points. (d) Standard deviation of $\Delta M_z$ at 36 measurement points on both the XZ plane and XY

plane for different $\theta_{mis}$.

We discuss why the stress distribution was narrow at high $\theta_{mis}$. Previous studies have reported that stress was caused in the diamond film around dislocations [32–34]. We evaluated the number of dislocations in the substrate and CVD film based on the number of etch pits created by exposing the surfaces of the diamonds to H$_2$ and O$_2$ plasma [35]. In H$_2$ and O$_2$ plasmas, dislocations are a source of fast dissolution, and etch pits form where the dislocations exist. First, a 2 mm square (111) diamond substrate was cut into four pieces. One was exposed to H$_2$ and O$_2$ plasma for 3 min, and the etch pits were counted using an atomic force microscope (AFM). Second, a CVD film with a thickness of 30 μm was synthesized on another piece of the substrate, and the number of etch pits in the CVD film was counted similarly. To generate H$_2$ and O$_2$ plasmas, hydrogen and oxygen gases were flowed at 100 and 2 sccm, respectively. The pressure, microwave power, and temperature of the diamond surface were set to 20 kPa, 3.5 kW, and 850 °C, respectively. In practice, the number of dislocations was evaluated for substrates with $\theta_{mis}$ of 3.3° and 10°. Some etch pits caused by dislocations are circled in red in Fig.4. The number of etch pits on the substrate for both $\theta_{mis}$ were almost the same value of approximately 10 as shown in Fig.4 (a) and (b). However, the number of etch pits on the CVD film was approximately 20 for $\theta_{mis}=$ 3.3° (Fig.4 (c)), whereas etch pits could not be identified for $\theta_{mis} = $ 10° (Fig.4 (d)). These results suggest that dislocations were less likely to form in the CVD films with high $\theta_{mis}$. This seems to be related to the observation that the lateral growth of the CVD film was accompanied with high steps at high $\theta_{mis}$ as shown in Fig.4 (d). For 4H-SiC growth[36–38], it was reported that when step-flow growth occurred with higher steps, threading dislocations from the substrate were transformed into other defects. Assuming an atomically flat diamond surface, the terrace width ($w$) is calculated as $w = d/\tan(\theta_{mis})$ where $d$=0.21 nm is the single-step height on the (111) surface [39]. Hence, the terrace width is about 3 times smaller for $\theta_{mis}=$ 10° than for $\theta_{mis} = $ 3.3°. In other words, individual atomic steps become closer to each other at higher $\theta_{mis}$. Therefore, step bunching is more likely to be induced at higher $\theta_{mis}$, and higher steps are more likely to be formed. Thus, we propose that the formation of higher steps due to the increased $\theta_{mis}$ have contributed to the reduction of dislocations in the CVD diamond films, resulting in reduction in the stress distribution. To clarify the mechanism of stress-distribution relaxation at high $\theta_{mis}$, it would be effective to further analyze the spatial distribution of dislocations in the CVD film using TEM and X-ray topography.

In this study, the diamond film contained approximately 20–30 ppm of $N_s^0$ and 1.1% of $^{13}C$. On the other hand, the relaxation of stress would be more effective for diamond films with lower $N_s^0$ concentrations. Applying the stress-mitigation technique presented in this study to diamond films with lower $N_s^0$ and $^{13}C$ densities would offer a significant advantage for extending $T_2^*$.

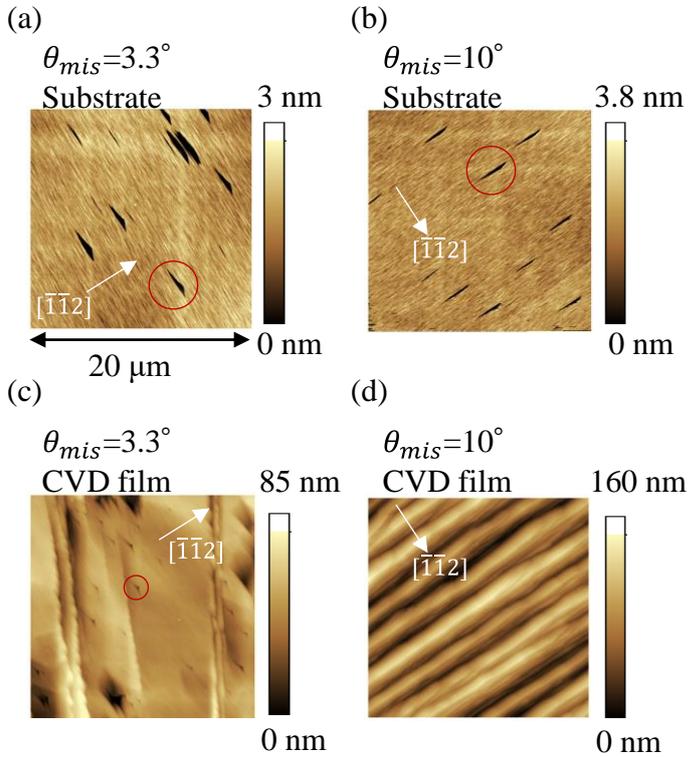

Fig. 4 (Color online) (a) and (b) AFM images of (111) diamond substrate surface with $\theta_{mis}$ of 3.3° and 10° after exposure to $H_2$ and $O_2$ plasma, respectively. (c) and (d) AFM images of the CVD film surfaces with $\theta_{mis}$ = 3.3° and 10°, respectively, after exposure to $H_2$ and $O_2$ plasma. Some of the etch pits are circled in red.

## 4. Conclusion

We extended the $T_2^*$ of large-ensemble NV centers by mitigating the stress distribution in thick diamond films with perfectly aligned NV centers. Measurements of $T_2^*$ with a large-excitation volume system showed that $T_2^*$ increased to approach the value limited only by the spin bath with increasing $\theta_{mis}$. Microscopic stress imaging using a confocal microscope showed that the stress accumulated along the distance from the substrate. In addition, the standard deviations of $\Delta M_z$ decreased by factors of approximately 24 on the cross section (XZ plane) and 11 on the surface (XY plane) of the CVD film as $\theta_{mis}$ increased from 2° to 10°. The suppression of dislocation formation in the CVD film at high $\theta_{mis}$ may have contributed to the reduced stress distribution. Applying the stress-mitigation technique presented in this study to diamond films with lower $N_s^0$ and $^{13}C$ densities would offer a significant advantage for extending $T_2^*$. The method developed in this study represents an important step in the preparation of high-quality diamonds for highly sensitive quantum sensors.

Declaration of competing interest

The authors declare that they have no known competing financial interests or personal relationships that could have appeared to influence the work reported in this paper.

Acknowledgements

This work was supported by MEXT Quantum Leap Flagship Program (MEXT Q-LEAP Grant No. JPMXS0118067395), Nanotechnology Platform Program of MEXT, Japan (Grant No. JPMXP09F-21- IT-014), JST SPRING (Grant No. JPMJSP2106) and JSPS KAKENHI (Grant No. 20H00355).

**Supporting Information for "Extending spin dephasing time of ensemble Nitrogen-Vacancy centers by mitigating stress distribution on highly-misoriented chemical vapor deposition diamond"**

T. Tsuji[1], T. Sekiguchi[1], T.Iwasaki[1] and M.Hatano[1*]

*[1] Department of Electrical and Electronic Engineering, School of Engineering,*

*Tokyo Institute of Technology, 2-12-1 Ookayama, Meguro, Tokyo 152-8552,*

*Japan*

A. Experimental setup for measuring the $T_2^*$ with a large excitation volume.

Fig.S1 (a) shows a schematic of the experimental setup for the large-excitation volume system. A laser focused on 20 µm diameters was irradiated on the diamond sample. The fluorescence of the NV centers was detected using a photodiode. A ring magnet was placed behind the coplanar waveguide. The $T_2^*$ of the NV center was reduced by the magnetic field inhomogeneity in the detection volume [S1]. Therefore, to suppress the magnetic field inhomogeneity, we used a ring-shaped magnet with an outer diameter, inner diameter, height, and residual magnetization of 42 mm, 34 mm, 6 mm, and 1030 mT, respectively. The left side of Fig.S1 (b) shows the z-dependence of the magnetic field intensity in the z-axis direction, which was calculated using *Magpylib* (https://magpylib.readthedocs.io/en/latest/index.html). The right side of Fig.S4 (b) shows a zoomed in view of the 19–22 µm range of the z in the left figure. The diamond sample was placed at approximately z=20.4 mm, where the magnetic-field deviation was the smallest. When the stress distribution was measured using confocal microscopy, the diamond was tilted, as described in Section C. The coplanar waveguide with the diamond sample was also tilted so that the laser could irradiate the region where the stress distribution was measured using the confocal microscope.

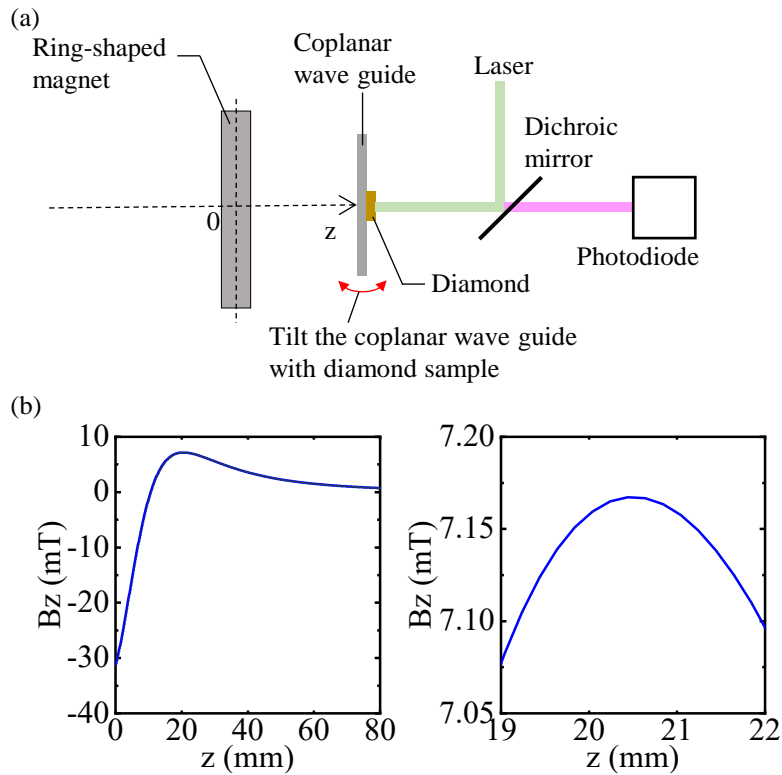

Fig.S1 (a) Schematic of the experimental of the large excitation volume system. (b) z dependence of the magnetic field intensity in the z-axis direction.

## B. Details of measuring principle for the stress distribution

As shown in Eq.(7) (main text), we can calculate

$$\alpha(P) - \alpha(P_0) =$$

$$\approx \Delta M_z(P) + \frac{3}{2}\gamma^2(B_x^2 + B_y^2)\left(\frac{1}{D + M_z(P)} - \frac{1}{D + M_z(P_0)}\right) \quad (\because D \gg \gamma B_z)$$

$$\approx \Delta M_z(P) - 3/2\gamma^2(B_x^2 + B_y^2)\left(\frac{M_z(P) - M_z(P_0)}{D^2}\right) \quad (\because D \gg M_z)$$

$$= \Delta M_z(P)\left(1 - \frac{3\gamma^2(B_x^2 + B_y^2)}{2D^2}\right),$$

where the deviation of $M_z$ is defined as $M_z(P) - M_z(P_0) = \Delta M_z(P)$. To reduce $\frac{3\gamma^2(B_x^2+B_y^2)}{2D^2}$, a two-axis goniometer was introduced to adjust the orientation of the diamond substrate and direction of the magnetic field. Consequently, the direction of the magnetic field, which had a magnitude of approximately 4 mT, was set to within $\cong 5°$ of the [111] direction. (The details are given in section C.) In this configuration, we can calculate $\frac{3/2\gamma^2(B_x^2+B_y^2)}{D^2} < 1 \times 10^{-4}$.

## C. Experimental setup for measuring the stress distribution.

Fig.S2 (a) shows a schematic of the experimental setup for measuring the stress distribution using a confocal microscope. A ring-shaped magnet was placed around the objective lens so that the detection point was located near the central axis of the magnet. As shown in Fig.S2 (b), to reduce the $\frac{3\gamma^2(B_x^2+B_y^2)}{2D^2}$ described in section B, the orientation of the diamond sample was controlled so that the direction of the magnetic field aligned with the [111] orientation of the diamond sample, which was measured using XRD. Consequently, the direction of the magnetic field was set to within $\cong 5°$ of the [111] direction.

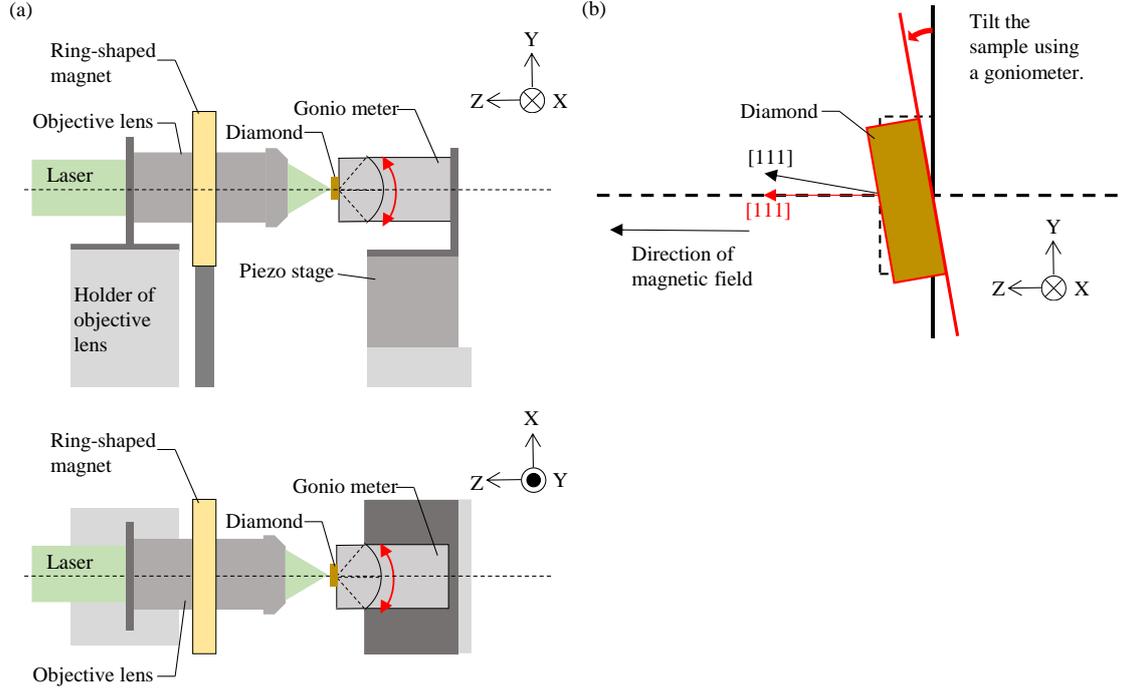

Fig.S2 (a) Schematic of the experimental setup for measuring the stress distribution with the confocal microscope. (b) Illustration of a diamond substrate tilted using a goniometer to align the direction of the magnetic field with the [111] direction of the diamond.

### D. Measuring $T_2$ using confocal microscope for calculating $[N_s^0]$ to obtain $T_2^*$ {Electron spin}.

To obtain $T_2^*$ {Electron spin}, we calculated the $[N_s^0]$ in the CVD film using Eq.(3) (Main text). $[N_s^0]$ in the CVD film was calculated from $T_2$ measured by the Hahn echo sequence using the confocal microscope. $T_2$ was measured at 3 μm intervals along the depth direction of the CVD film (the Z direction), and the average value of these $T_2$ was used for calculating $[N_s^0]$. Fig. S3 (a) shows the spin-echo signal of the CVD film with $\theta_{mis}$ of 10° at Z= 0 μm, which was fitted to the expression of $A\exp\left(-\frac{\tau}{T_2}\right)^p$. The fitting results showed that $T_2$=6.4 μs and the fitting error of $T_2$ was 0.2 μs. Fig. S3 (b) shows the z-direction dependence of $T_2$ for different $\theta_{mis}$. Table.1 (main text) shows the average value of $T_2$, $[N_s^0]$ and $T_2^*$ {Electron spin} for all $\theta_{mis}$. $[N_s^0]$ and $T_2^*$ {Electron spin} were calculated using Eqs. (2) and (3), respectively.

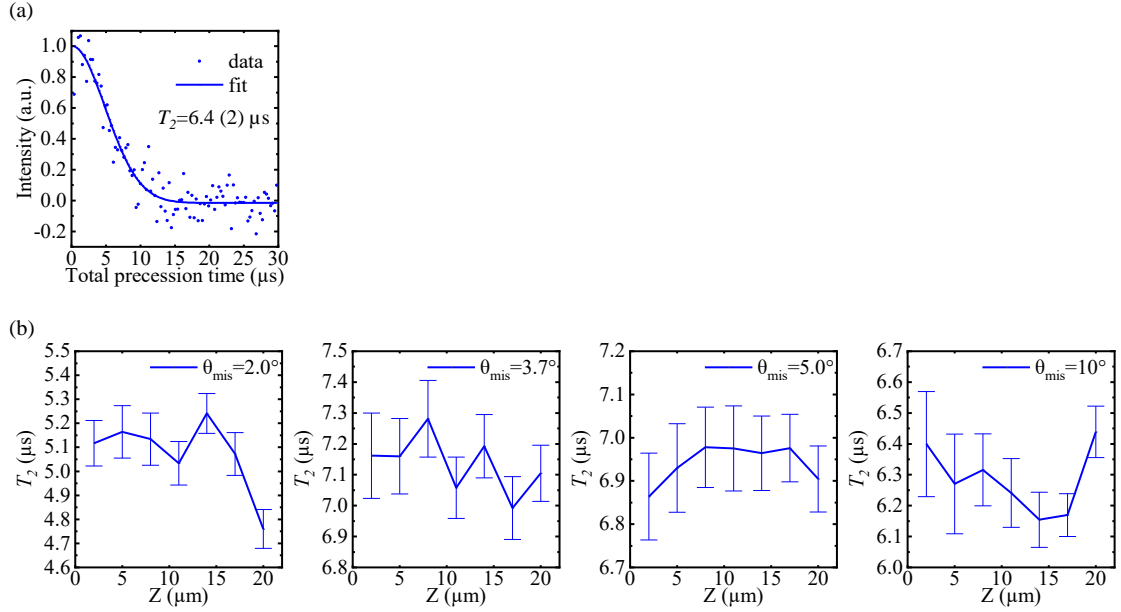

Fig.S3 (a) Spin-echo signal of the CVD film with $\theta_{mis}$ of 10° at Z= 0 µm. (b) Z-direction dependence of $T_2$ for different $\theta_{mis}$. Error bars corresponded to the fitting errors.

E. Measurement accuracy of $\Delta M_z$

We measured the two resonance frequencies ($f_{\pm}(P)$) of the NV center at each measurement position $P$ using the Ramsey sequence. First, we applied microwaves ($f_{mw}$) to obtain the Ramsey fringes. Fig.S4 shows the Ramsey fringes of the NV centers in the CVD film with $\theta_{mis}$ of 10°. The Ramsey fringes were fitted using the following function

$\exp\left(-\frac{\tau}{T_2^*}\right)[a_1 \cos(2\pi(f_0 - f_h)\tau + \varphi_1) + a_2 \cos(2\pi f_0 \tau + \varphi_2) + a_3 \cos(2\pi(f_0 + f_h)\tau + \varphi_3)]$,

where $f_h$ represents the hyperfine splitting. In this case, the resonance frequency of the NV center is expressed as $f_+ = f_{mw} - f_0$. This procedure was performed for two resonance frequencies at the NV center ($f_{\pm}$). The measurement accuracy of $f_0$ was approximately 13 kHz, with an integration time of 10 min. Therefore, the accuracy of $\Delta M_z (\approx \alpha(P) - \alpha(P_0) = \frac{f_-(P)+f_+(P)}{2} - \frac{f_-(P_0)+f_+(P_0)}{2})$ was approximately 13 kHz.

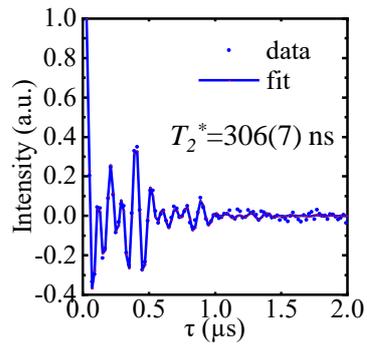

Fig.S4 Ramsey fringe of NV centers in CVD film with $\theta_{mis}$ of 10°. The fitting results showed that $T_2^*$=306 ns, $f_0$=8.057 MHz, and the fitting error of $f_0$ was 13 kHz.